\documentclass[10pt,conference]{IEEEtran}
\ifCLASSINFOpdf
\else
\fi

\usepackage{times}
\usepackage{latexsym}
\usepackage{stfloats}
\usepackage{citesort}
\usepackage{graphicx, amsfonts,amsmath, amssymb,array,url,stfloats}
\usepackage[noadjust]{cite}
\usepackage{graphicx}
\usepackage{subfigure}
\usepackage{enumerate}
\usepackage{hyperref}

\newtheorem{theorem}{Theorem}

\newtheorem{lemma}{Lemma}

\begin{document}
\title{Energy Efficiency Optimization in Hardware-Constrained Large-Scale MIMO Systems}

\author{\IEEEauthorblockN{Xinlin~Zhang\IEEEauthorrefmark{1}, Michail~Matthaiou\IEEEauthorrefmark{2}\IEEEauthorrefmark{1}, Mikael Coldrey\IEEEauthorrefmark{3},  and Emil~Bj\"ornson\IEEEauthorrefmark{4}}
\IEEEauthorblockA{\IEEEauthorrefmark{1}Department of Signals and Systems, Chalmers University of Technology, Gothenburg, Sweden}
\IEEEauthorblockA{\IEEEauthorrefmark{2}School of Electronics, Electrical Engineering and Computer Science, {Queen's} University Belfast, Belfast, U.K.}
\IEEEauthorblockA{\IEEEauthorrefmark{3}Ericsson Research, Ericsson AB, Gothenburg, Sweden}
\IEEEauthorblockA{\IEEEauthorrefmark{4}Department of Electrical Engineering (ISY), Link\"{o}ping University, Link\"{o}ping, Sweden}
%\IEEEauthorblockA{\IEEEauthorrefmark{5}Department of Signal Processing, ACCESS, KTH Royal Institute of Technology, Stockholm, Sweden}

E-mail: xinlin@chalmers.se, m.matthaiou@qub.ac.uk, mikael.coldrey@ericsson.com,  emil.bjornson@liu.se}


\IEEEoverridecommandlockouts
\IEEEpubid{\makebox[\columnwidth]{978-1-4799-5863-4/14/\$31.00~\copyright~2014 IEEE \hfill} \hspace{\columnsep}\makebox[\columnwidth]{ }} \maketitle
%\maketitle

\begin{abstract}
Large-scale multiple-input multiple-output (MIMO) communication systems can bring substantial improvement in spectral efficiency and/or energy efficiency, due to the excessive degrees-of-freedom and huge array gain. However, large-scale MIMO is expected to deploy lower-cost radio frequency (RF) components, which are particularly prone to hardware impairments. Unfortunately, compensation schemes are not able to remove the impact of hardware impairments completely, such that a certain amount of residual impairments always exists. In this paper, we investigate the impact of residual transmit RF impairments (RTRI) on the spectral and energy efficiency of training-based point-to-point large-scale MIMO systems, and seek to determine the optimal training length and number of antennas which maximize the energy efficiency. We derive deterministic equivalents of the signal-to-noise-and-interference ratio (SINR) with zero-forcing (ZF) receivers, as well as  the corresponding spectral and energy efficiency, which are shown to be accurate even for small number of antennas. Through an iterative sequential optimization, we find that the optimal training length of systems with RTRI can be smaller compared to ideal hardware systems in the moderate SNR regime, while larger in the high SNR regime. Moreover, it is observed that RTRI can significantly decrease the optimal number of transmit and receive antennas. 
\end{abstract}


% \fi
\IEEEpeerreviewmaketitle

\section{Introduction}\label{Sec:Intro}
Large-scale MIMO is regarded as a key enabler to boost the performance of future wireless communication networks \cite{Marzetta2010Massive_TWCOM}. This is achieved by deploying unconventionally large number of transmit and/or receive antennas and by exploiting channel reciprocity in time-division duplex (TDD) operation. It has been shown that extra antennas can greatly increase the spectral efficiency and, up to a certain limit, the energy efficiency \cite{Hien2013Massive_TCOM,Emil2014EE_TWCOM}. Moreover, large-scale MIMO systems are more robust compared to conventional MIMO systems, in the sense that they are more resilient to small-scale fading, inter-user interference, and to some extent, hardware impairments \cite{emil2013est}. 

With large-scale MIMO, the hardware accuracy constraints can be relaxed \cite{emil2013est}; thus, these systems can deploy lower-quality RF components which are particularly prone to hardware impairments (e.g., I/Q imbalance, amplifier non-linearities, and phase noise). The impact of such individual hardware impairments is usually mitigated by using analog and digital signal processing algorithms \cite{schenk2008rf}. However, these techniques are not able to completely remove hardware impairments, such that a certain amount of residual distortions always remains.  
These residual impairments stem from, for example, time-varying hardware characteristics which cannot be accurately parameterized and estimated, as well as, the randomness induced by different types of noise and imperfect compensation schemes. The impact of residual hardware impairments has been only scarcely investigated in few works recently. The authors in \cite{schenk2008rf, studer2010residual}  characterized and verified experimentally that the distortion caused by the impairments behaves as additive and independent Gaussian noise.
This Gaussian behavior can be interpreted by the law of large numbers, when the residual distortions from many independent and different sources add up together. In \cite{Emil2013imp_COML, xinlin2014capacity_ICC}, the authors respectively showed that impairments fundamentally limit the MIMO channel capacity in the high signal-to-noise (SNR) regime, while for low transmit power such impact is negligible; however, these results are based on the assumption of perfect channel state information (CSI). In \cite{emil2013est} the authors investigated large-scale MIMO systems with imperfect CSI, and showed that residual hardware impairments imposed an estimation error floor regardless of the SNR values; nonetheless, it was shown that large-scale MIMO systems can still achieve relatively high spectral efficiency and energy efficiency. However, they did not pursue any resource allocation analysis, which is of pivotal importance for the performance optimization of training-based MIMO systems.

In this paper, we consider a training-based point-to-point large-scale MIMO system with RTRI. We focus on two system performance metrics, namely spectral efficiency and energy efficiency. In particular, we derive $\textit{deterministic equivalents}$ of the spectral efficiency and energy efficiency for systems with zero-forcing (ZF) receivers. These deterministic expressions are exact in the large system limit, but are accurate also for ``not so large'' systems. To maximize the energy efficiency, we perform an iterative sequential optimization of the training length, and the number of transmit and receive antennas. We show that RTRI have significant impact on the optimal values of training length, as well as the number of antenna elements. This work provides guidelines on designing these parameters in large-scale MIMO systems with hardware impairments.

\textit{Notation:} Upper and lower case boldface letters denote matrices and vectors, respectively. The trace of a matrix is expressed by $\mathrm{tr}\left(\cdot\right)$. The $n \times n$ identity matrix is represented by $\mathbf I_n$. The expectation operation is $\mathbb {E} [\cdot]$. The superscripts $(\cdot)^ H$, $(\cdot)^{-1}$ and $(\cdot)^{\dagger}$ stand for Hermitian transposition, matrix inverse and pseudo-inverse, respectively. The Frobenius norm and  spectral norm are denoted by $\left\|\cdot\right\|_F$ and $\left\|\cdot\right\|_2$, respectively. The symbol $\mathcal{CN}\left(\mathbf m, \boldsymbol\Sigma\right)$ denotes a circularly-symmetric complex Gaussian (CSCG) distribution with mean $\mathbf m$ and covariance $\boldsymbol\Sigma$. For any matrix $\mathbf H\in\mathbb C^{m\times n}$, $\mathbf h_i$ is the $i$-th column of $\mathbf H$.

\section{Signal and System Models}
In this paper, we consider a point-to-point link with a block fading channel of coherence length $T$. During each block, the channel $\mathbf H\in\mathbb C^{N_r\times N_t}$ is constant, and is a realization of the uncorrected Rayleigh fading model, where $N_t$ and $N_r$ are the numbers of transmit and receive antennas, respectively. %Channel realizations between different blocks are assumed to be independent.

Each channel coherence block of length $T$ is split into a channel training phase and a data transmission phase. A number of $T_p$ channel uses (c.u.s) is devoted to channel training, while the rest $T_d\triangleq T-T_p$ c.u.s are reserved for data transmission.  During the training phase, each transmit antenna transmits known orthogonal pilot sequences to the receiver. However, due to the existence of transmit hardware impairments, there will be a mismatch between the intended training sequences and the actual transmitted signal. As a result, the receiver estimates the current channel realization based on a perturbed observation
\begin{equation}\label{eq:TrainingModel}
\mathbf Y_p = \sqrt{\frac{\rho}{N_t}}\mathbf H \left(  \mathbf S_p\! +\! \boldsymbol\Delta_p \right) + \mathbf V_p, ~\mathrm{tr}\left(\mathbf S_p\mathbf S_p^H\right) = N_tT_p,
\end{equation}
where  $\mathbf S_p\in\mathbb C^{N_t\times T_p}$  is the deterministic matrix of training sequences, which is known by the receiver, $\mathbf Y_p$ is the $N_r\times T_p$ received matrix, and $\mathbf V_p$ is the receiver noise. Each element of $\mathbf H$ and $\mathbf V_p$ follows a $\mathcal{CN}(0,1)$ distribution. The average SNR at each receive antenna is denoted by $\rho$. The distortion caused by transmit hardware impairments is introduced by $\boldsymbol{\Delta}_p\in\mathbb C^{N_t\times T_p}$. From the discussion in Section \ref{Sec:Intro} and \cite{schenk2008rf, studer2010residual}, it is known that $\boldsymbol{\Delta}_p$ can be well characterized by 
the following Gaussian model
\begin{align}
{\boldsymbol\delta_p}_{i}&\sim \mathcal{CN}\left(\mathbf 0, \delta^2\mathbf I_{N_t}\right), \mathbb E\left[{\boldsymbol\delta_p}_{i}{\boldsymbol\delta_p}_{j}^H\right] = \mathbf 0,\\  ~&\textrm{for} ~i,j =1,2,\dots,T_p,i\neq j,\notag
\end{align}
where the proportionality parameter $\delta$ quantifies the level of residual transmit hardware impairments. Note that $\delta$ appears in practical applications as the error vector magnitude (EVM) \cite{holma2011LTE}, which is commonly used to measure the quality of RF transceivers. For instance, 3GPP LTE has EVM requirements in the range $\left[0.08,0.175 \right]$ \cite{holma2011LTE}, where high spectral efficiency requires smaller EVMs. The relationship between $\delta$ and EVM is defined as
\begin{equation}
\mathrm{EVM} \triangleq \sqrt{\frac{\mathbb{E}_{\boldsymbol\Delta_p}\left[ \left\| \boldsymbol\Delta_p \right\|_F^2 \right]}{\mathbb{E}_{\mathbf S_p}\left[ \left\| \mathbf S_p\right\|_F^2 \right]}} = \delta.
\end{equation}
Evidently, when $\delta\! = \!0$, the system model simplifies to the ideal hardware case. From \cite{xinlin2014training_ICC}, we know that the linear minimum mean-squared error (LMMSE) channel estimator is given by 
\vspace{-0.1cm}
\begin{equation}\label{eq:estimator}
\hat{\mathbf H} = \mathbf Y_p\bigg( \mathbf S_p^H{\mathbf S_p} + \left( \delta^2\rho + 1 \right)\mathbf I_{T_p} \bigg)^{-1}\mathbf S_p^H, 
\end{equation}
with $\mathbf S_p\mathbf S_p^H = T_p\mathbf I_{N_t}$. It is also known that each element of $\hat{\mathbf H}$ has variance $\sigma^2_{\hat{\mathbf H}} = \frac{\epsilon}{1+\epsilon}$, where $\epsilon \triangleq \frac{\rho T_p}{N_t(\rho\delta^2+1)}$. According to the orthogonality principle of LMMSE estimators, the variance of each element of the estimation error $\tilde{\mathbf H}$ equals  $\sigma^2_{\tilde{\mathbf H}} = 1-\sigma^2_{\hat{\mathbf H}} = \frac{1}{1+\epsilon}$. For any given $T_p$ and $N_t$, as $\rho\rightarrow\infty$, $\sigma^2_{\tilde{\mathbf H}}\rightarrow 1/\left(1+\frac{T_p}{N_t\delta^2}\right)$, which, in contrast to the case with ideal hardware, means that the variance of the estimation error converges to a non-zero value, thus an estimation error floor occurs; yet, we can still improve the estimation accuracy by increasing $T_p$.

During the data transmission phase, RTRI affect the system as follows:
\begin{align}\label{eq:DataModel}
\mathbf Y_d \!= \!\sqrt{\frac{\rho}{N_t}}\mathbf H \left( \mathbf S_d \!+\! \boldsymbol\Delta_d\right)\! +\! \mathbf V_d,
~\mathbb{E}\Big[\mathrm{tr}\left(\mathbf {S}_d^H\mathbf S_d\right)\Big]\! =\! N_t T_d,
\end{align}
where $\mathbf S_d \in \mathbb{C}^{N_t \times T_d}$ is the random matrix of data symbols with each entry following a $\mathcal{CN}(0,1)$ distribution, and $\mathbf Y_d$ is the $N_r \times T_d$ received signal matrix. The distortion noise caused by transmit hardware impairments is characterized as 
\begin{align}
{\boldsymbol\delta_d}_{i}\sim \mathcal{CN}\left(\mathbf 0, \delta^2\mathbf I_{N_t}\right), \mathbb E\left[{\boldsymbol\delta_d}_{i}{\boldsymbol\delta_d}_{j}^H\right] = \mathbf 0,\\~\textrm{for}~i,j=1,2,\dots,T_d,i\neq j.
\end{align}

\section{Spectral efficiency analysis}
In this section, we analyze the spectral efficiency of the proposed system model with ZF receivers. Each column of the received signal $\mathbf y_d$ in (\ref{eq:DataModel}) can be written as follows
\begin{align}
{{\mathbf y}_d} & = \sqrt{\frac{\rho}{N_t}}\left({\hat {\mathbf H}}{\mathbf s_d} + {\tilde {\mathbf H}}{\mathbf s_d} + \left( {\hat {\mathbf H}} + {\tilde {\mathbf H}}\right){\boldsymbol\delta_d}\right) +\mathbf v_d\\
& = \!\sqrt{\frac{\rho}{N_t}}\hat {\mathbf h}_k{s_d}_k \!\!+ \!\! \underbrace{\sqrt{\frac{\rho}{N_t}}\!\!\left( \sum_{\substack{ i = 1 \\ i \neq k }}^{N_t}\!\hat {\mathbf h}_i{s_d}_i\!\! +\!\! {\tilde {\mathbf H}}{\mathbf s_d}\! +\! \left( \!{\hat {\mathbf H}} \!+\! {\tilde {\mathbf H}}\!\right){\boldsymbol\delta_d}\!\!\right)\!\! +\!\mathbf v_d}_{\triangleq{\mathbf z_d}_k}, \notag
\end{align}
%\&~~~~~~~~~~~~~~~~~~~~~~~~~~~~~~~~~~~~~~~~k = 1, 2, \dots, N_t,\notag
for $k = 1, 2, \dots, N_t$, where ${s_d}_k$ is the transmitted data symbol from the $k$-th transmit antenna, and ${\mathbf z_d}_k$ is the total effective noise plus interference on the $k$-th spatial stream. To recover the signal ${s_d}_k$, ${{\mathbf y}_d}$ is multiplied by the $k$th row of ${{\hat{\mathbf H}}^{\dagger}}$. Since $\hat{\mathbf H}^H$ is known at the receiver, we can express the SINR of ZF receivers on the $k$-th spatial stream as 
\begin{align}
\gamma_{k,\mathrm{ZF}} &= \frac{1}{\delta^2 + c_0\left[\left(\bar {\mathbf H}^H\bar {\mathbf H}\right)^{-1}\right]_{k,k}},\label{eq:SINR_ZF}
\end{align}
where we have defined $c_0 \triangleq \frac{N_t\left(\rho+\rho\delta^2+1+\epsilon\right)}{\rho \epsilon}$ for notational convenience. We also define $\bar{\mathbf H} \triangleq \frac{1}{\sigma^2_{\hat{\mathbf H}}}\hat{\mathbf H}$, which approximately has uncorrelated  $\mathcal{CN}(0,1)$ entries \cite{xinlin2014training_ICC}.
Then, the ergodic spectral efficiency can be evaluated as 
\vspace{-0.1cm}
\begin{equation}\label{eq:Rate}
R_\mathrm{ZF}(\gamma) =  \frac{T_d}{T}\mathbb E\left[ \sum_{k=1}^{N_t}\log_2\left(1+\gamma_{k, \mathrm{ZF}}\right)\right].
\end{equation}
For large-scale MIMO systems, we can derive deterministic equivalents of (\ref{eq:Rate}) with the help of the following lemma:
\begin{lemma}\label{lemma:TraceLemma}{\textit{(\cite[Corollary 6.3]{Couillet2011a})}}
Let $\mathbf X$ be composed of $n<N$ columns of an $N\times N$ Haar matrix and $\mathbf x$ is an arbitrary column of $\mathbf X$, and let $\mathbf A\in\mathbb C^{N\times N}$ be independent of $\mathbf X$ and have uniformly bounded spectral norm. Then, as $N$ grows large, we have:\vspace{-0.3cm} 
\begin{align}
\mathbf x^H\mathbf A\mathbf x-\frac{1}{N}\mathrm{tr}\left(\mathbf A\right)\stackrel{a.s.}\longrightarrow 0,
\end{align}
\end{lemma}\vspace{-0.2cm} 
where $\stackrel{a.s.}\longrightarrow$ denotes almost sure convergence.

We now give the deterministic equivalent of the SINR of ZF receivers in the following theorem:
\begin{theorem}\label{theorem:bothlarge}
As $N_r, N_t \rightarrow\infty$, with a finite ratio $\beta = \frac{N_r}{N_t}\in(1, \infty)$\footnote{Due to page limit, we do not consider the case $\beta = 1$ ($N_r = N_t$). One can find a detailed study of this case, for example, in \cite{Eldar2003ZFAsym_TIT}.}, the SINRs of ZF receivers, converge almost surely to the right-hand side deterministic equivalent
\begin{align}
\gamma_{\mathrm{ZF}}^{N_r,N_t\rightarrow\infty} &\xrightarrow{a.s.} \bar{\gamma}_{\mathrm{ZF}} \triangleq \frac{N_r-N_t}{\delta^2N_r+(c_1-\delta^2)N_t}\label{eq:deter_ZF},
\end{align}
where $c_1 \triangleq \frac{\rho+\rho\delta^2+1+{\epsilon}}{\rho{\epsilon}}$.
\end{theorem} 
\begin{IEEEproof}
The singular value decomposition (SVD) of $\bar{\mathbf H}$ is denoted by $\bar{\mathbf H} = \mathbf U\boldsymbol\Sigma\mathbf V^H$, where $\mathbf U$ and $\mathbf V$ are unitary matrices, while $\boldsymbol \Sigma$ is a $N_r\times N_t$ diagonal matrix with the diagonal elements containing the singular values of $\bar{\mathbf H}$. Consequently, we have
\vspace{-0.2cm}
\begin{align}
\left[\left(\bar {\mathbf H}^H\bar {\mathbf H}\right)^{-1}\right]_{k,k} \!&= \left[\mathbf V\boldsymbol\Lambda\mathbf V^H\right]_{k,k}\notag\\ 
&= \mathbf v_k^H\boldsymbol\Lambda\mathbf v_k \underset{(a)}{\stackrel{a.s.}\longrightarrow} \frac{1}{N_t}\textrm{tr}\left(\boldsymbol\Lambda\right) \stackrel{a.s.}\longrightarrow \frac{1}{N_r\!-\!N_t},\notag
\end{align}
where $\boldsymbol\Lambda \triangleq \left(\boldsymbol\Sigma^H\boldsymbol\Sigma\right)^{-1}$. Note that $\mathbf v_k$ is known to satisfy the conditions of $\mathbf x$ in Lemma \ref{lemma:TraceLemma} \cite{Eldar2003ZFAsym_TIT}, therefore $(a)$ follows readily. Substituting the above result into the SINR expression with ZF in (\ref{eq:SINR_ZF}) completes the proof.
\end{IEEEproof} 
%is a diagonal matrix with its diagonal elements being the eigenvalues of $\left(\bar {\mathbf H}^H\bar {\mathbf H}\right)^{-1}$

Theorem \ref{theorem:bothlarge} yields a deterministic equivalent of the SINR with ZF in the large-antenna regime, which does not rely on random channel realizations, as long as the ratio $\beta$ is fixed. This result is quite general and tractable and, as we will show later, is a very accurate approximation even for small number of antennas. Note that some practical interesting system setups, for instance, $N_t \ll N_r$ (usually appearing in the multi-user massive MIMO uplink), which corresponds to $\beta\rightarrow\infty$, can be treated as a special case of Theorem \ref{theorem:bothlarge}. 

%In particular, we will give this result in the following corollary
%\begin{corollary}\label{corollary:Nr_Large}
%For the case $N_t\ll N_r$, such that $\beta\rightarrow\infty$, the SINRs of ZF, MRC, and MMSE receivers, converge almost surely to the same deterministic equivalent\begin{equation}\label{eq:Nr_Large}
%\gamma^{\beta\rightarrow\infty} \stackrel{a.s.}\longrightarrow \bar{\gamma} = \frac{1}{\delta^2}.
%\end{equation}
%\end{corollary}
%From the above equation, we can see that the SINR reaches a limit when $N_t\ll N_r$, which is only quantified by the level of hardware impairments. In this case, increasing the transmit power does not provide any SINR gain. 

\begin{figure}[h!]
\begin{centering}
\includegraphics [height=0.26\textwidth]{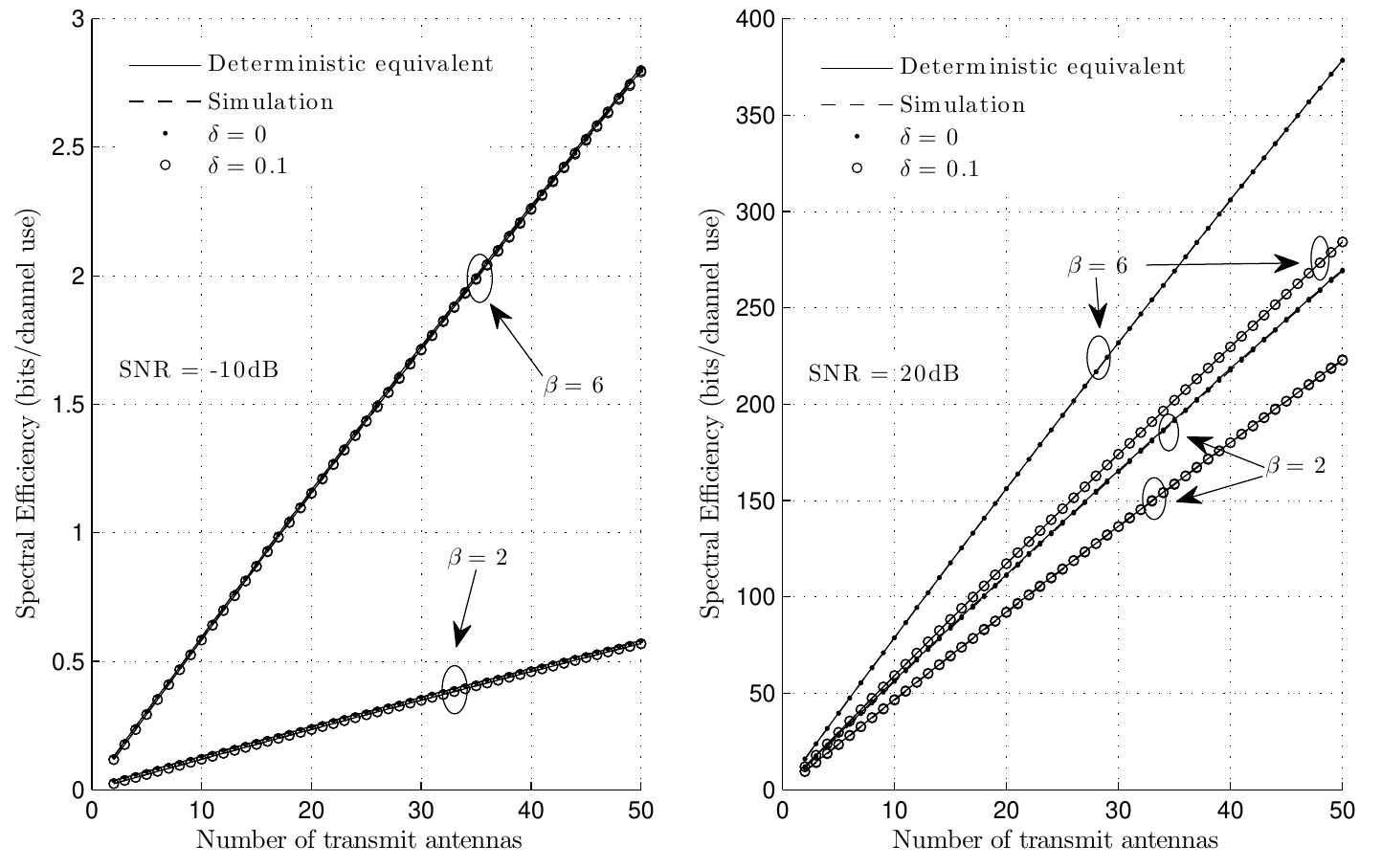}
\caption{Spectral efficiency at different SNR values.}\label{fig:SE}
\end{centering}
\end{figure}

With the asymptotic SINRs, we proceed to derive the spectral efficiency. 
Using continuous mapping theorem  \cite[Theorem 2.3]{Van2000Asymptotic} and dominated convergence theorem \cite[Theorem 16.4]{Billingsley2008Probability}, we can show that the spectral efficiency with ZF receivers converges to the right-hand side deterministic equivalent
\begin{equation}\label{eq:Rate_Asymp}
R^{N_r, N_t\rightarrow\infty}_\mathrm{ZF} \longrightarrow \bar{R}= \left(1-\frac{T_p}{T}\right)N_t\log_2\left(1+\bar{\gamma}_\mathrm{ZF}\right).
\end{equation}
Figure \ref{fig:SE} shows the spectral efficiency against the number of transmit antennas  for both low and high SNR values. Interestingly, (\ref{eq:Rate_Asymp}) is very accurate even for small number of antennas. At $\rho$ = 20dB, RTRI create an offset in spectral efficiency as compared to that of an ideal hardware system. The offset increases with the number of transmit antennas, as well as with the ratio $\beta$. Therefore, we can infer that RTRI significantly affect the spectral efficiency of high-rate large-scale MIMO systems. However, in the low SNR regime, as we can see from the left subplot, RTRI have only negligible impact.
  \vspace{-0.2cm}
\vspace{-0.5cm}
\section{Energy Efficiency Analysis}
Based on the results on spectral efficiency, we proceed to analyze the energy efficiency. Our main goal herein is to find the optimal training length, as well as the optimal antenna configuration that maximize the energy efficiency. The energy efficiency of a communication system is measured in bit/Joule and equals the ratio between the spectral efficiency and the total power consumption. In general, the power consumption comes from, for instance, generation of RF power, operating static circuits, and feeding RF components that are attached to each transceiver chain. Motivated by \cite{Auer2011EE_TWCOM}, we model the total power consumption (in Joule/c.u.) as 
\vspace{-0.1cm}
\begin{align}
P_\mathrm{total} &= N_tP_\mathrm{tx} + N_rP_\mathrm{rx} + P_0 + P_\mathrm{RF}/\eta,\notag
\end{align}
where $P_\mathrm{tx}$ and $P_\mathrm{rx}$ are the powers of RF components (including antennas, filters, converters, etc.) at the transmitter and receiver, respectively, $P_0$ is the static circuit power consumption, $P_\mathrm{RF}$ is the RF power generated by the power amplifier, while $\eta$ is the efficiency of the power amplifier. We now can define the energy efficiency for the case of ZF receivers as 
\vspace{-0.2cm}
\begin{equation}
\mathrm{EE} \triangleq \frac{R_\mathrm{ZF}}{P_\mathrm{total}}.
\end{equation}
Replacing $R_\mathrm{ZF}$ with its deterministic equivalent in (\ref{eq:Rate_Asymp}), we can well-approximate the energy efficiency as
\vspace{-0.2cm}
\begin{equation}\label{eq:EE_DE}
\mathrm{EE}\approx\overline{\mathrm{EE}}\triangleq\frac{\overline{R}}{P_\mathrm{total}}.
\end{equation}
For an arbitrary SNR value, $\overline{\mathrm{EE}}$ can be treated as a function of three parameters, $N_t$, $N_r$ and $T_p$. Intuitively, increasing a single value of the above parameters while keeping the other two fixed  does not  always increase the energy efficiency. For example, we can increase the spectral efficiency $\overline{R}$ by deploying more transmit antennas; however, in the meantime, transmit RF components consume more power, and the energy efficiency may even decrease. Ideally, we would like to determine the optimal parameters through solving the following optimization problem:
\vspace{-0.2cm}
\begin{align}
&\textrm{maximize}  ~~~\overline{\mathrm{EE}}(N_t,N_r,T_p)\notag\\
&\textrm{subject to}  ~~~N_t\leq T_p\leq T,\notag\\
&~~~~~~~~~~~~~~~1\leq N_t \leq T, \notag\\
&~~~~~~~~~~~~~~~N_r> N_t. \notag
\end{align}
However, the above problem is very difficult to solve, and we reformulate it to an equivalent form. Let $T_a \triangleq T_p - N_t$, and substitute this into 
$\overline{\mathrm{EE}}$, then we have
\begin{equation}\label{eq:EE_Deter}
\overline{\mathrm{EE}} \!=\! \frac{\left(1\!\!-\!\!\frac{N_t\!+\!T_a}{T}\right)\!N_t\!\log_2\!\left(\!1\!+\!\frac{\beta-1}{\delta^2(\beta\!-\!1)\!+\!\frac{N_t}{N_t\!+\!T_a}(\delta^2\!+\!1\!/\!\rho)(\delta^2\!+\!1\!/\!\rho\!+\!1)\!+\!1\!/\!\rho}\right)}{N_tP_\mathrm{tx} + \beta N_tP_\mathrm{rx} + P_0 + P_\mathrm{RF}/\eta}.
\end{equation}\vspace{-0.2cm} 
Therefore, the new optimization problem becomes
\begin{align}\label{eq:IterOptimization}
&\textrm{maximize}  ~~~\overline{\mathrm{EE}}(N_t,\beta,T_a)\\
&\textrm{subject to}  ~~~0\leq T_a\leq T-N_t,\notag\\
&~~~~~~~~~~~~~~~1\leq N_t\leq T, \notag\\
&~~~~~~~~~~~~~~~\beta> 1. \notag
\end{align} 
Unfortunately, it is still not tractable to perform a joint optimization to get the global optimum $(N_t^*, \beta^*, T_p^*)$. Nonetheless, we can use an iterative sequential optimization of $T_a$, $N_t$ and $\beta$, to find a set of suboptimal values which is very close to the global optimum. We will now establish the convexity of the objective function in (\ref{eq:IterOptimization}) with respect to each parameter.
\subsubsection{Optimal $T_p$} The composition rule in \cite[Exercise 3.32 (b)]{Boyd2004Convex} indicates that the product of a positive decreasing linear function and a positive increasing concave function is also concave. It follows readily that $\overline{\mathrm{EE}}$ is concave in $T_a$. Observing that $\overline{\mathrm{EE}}(T_a = 0) >0$ and $\overline{\mathrm{EE}}(T_a = N_t-T_p) = 0$, we can always find $T_a^*$ by
\vspace{-0.2cm}
\begin{equation}
T_a^* = \underset{T_a}\arg\left(\frac{\partial\overline{\mathrm{EE}}}{\partial T_a} = 0\right), ~\textrm{for}~0\leq T_a\leq T-N_t, \label{eq:Tp_iter}
\end{equation}
where $\frac{\partial\overline{\mathrm{EE}}}{\partial T_a}$ is the partial derivative of $\overline{\mathrm{EE}}$ with respect to $T_a$.
\subsubsection{Optimal $\beta$} The numerator of the objective function in (\ref{eq:EE_Deter}) is strictly concave in $\beta$, and the denominator is an affine function; thus, we can easily show that $\overline{\mathrm{EE}}$ is quasiconcave in $\beta$. Since $\overline{\mathrm{EE}}$ equals zero as $\beta\rightarrow 1$ and $\beta\rightarrow\infty$, $\beta^*$ can always can be found via
\begin{equation}
\beta^* = \underset{\beta}\arg\left(\frac{\partial\overline{\mathrm{EE}}}{\partial \beta} = 0\right), ~\textrm{for}~\beta > 1.\label{eq:beta_iter}
\end{equation}
\subsubsection{Optimal $N_t$} We cannot determine the convexity of $\overline{\mathrm{EE}}(N_t)$ in the entire feasible set. However, it is straightforward to prove that $\overline{\mathrm{EE}}(N_t)$ is quasiconcave for $N_t\leq \frac{T-T_a}{2}$, and monotonically decreasing for $N_t > \frac{T-T_a}{2}$. Therefore, the global optimum should satisfy the stationarity condition
\begin{equation}
N_t^* = \underset{N_t}\arg\left(\frac{\partial\overline{\mathrm{EE}}}{\partial N_t} = 0\right),~\mathrm{for}~1\leq N_t\leq\frac{T-T_a}{2}.\label{eq:Nt_iter}
\end{equation}

%More specifically, for different receivers, we want to find
%\begin{align}
%N_t^* &= \underset{N_t}\arg\left(\frac{\partial\bar{\mathrm{EE}}}{\partial N_t} = 0\right),\label{eq:Nt_iter}\\
%\beta^* &= \underset{\beta}\arg\left(\frac{\partial\bar{\mathrm{EE}}}{\partial \beta} = 0\right),\label{eq:beta_iter}\\
%T_p^* &= \underset{T_p}\arg\left(\frac{\partial\bar{\mathrm{EE}}}{\partial T_p} = 0\right).\label{eq:Tp_iter}
%\end{align}
Given an arbitrary initial feasible set $({T_p}_0, \beta_0, {N_t}_0)$, the proposed iterative algorithm is implemented as follows:
\begin{enumerate}
\item Update $T_p^*$ according to (\ref{eq:Tp_iter});
\item Update $\beta^*$ using (\ref{eq:beta_iter});
\item Update $N_t^*$ with the optimal value from (\ref{eq:Nt_iter});
\item Repeat 1)--3) until $\overline{\mathrm{EE}}$ converges.
\end{enumerate}
Each step in 1)--3) can be solved by implementing the bisection method. Convergence happens when the difference of $\overline{\mathrm{EE}}$ in two consecutive steps is smaller than the threshold $\overline{\mathrm{EE}}_\mathrm{th}$. This algorithm is guaranteed to converge since the finite $\overline{\mathrm{EE}}$ is nondecreasing with each parameter.
  \vspace{-0.3cm}
\begin{table}[h]
  \caption[Table caption text]{Simulation Parameters}
  \vspace{-0.1cm}
  \begin{tabular}{ |c | c || c | c|}
    \hline
    \textbf{Parameter} & \textbf{Value} & \textbf{Parameter} & \textbf{Value}\\ \hline
    Coherence bandwidth: $B$ & 180 kHz & $P_0$& 2 W$\cdot$S \\
    Coherence time: $T$ & 32ms $\cdot$ $B$ $\cdot$ c.u. &$P_\mathrm{tx}$&1 W$\cdot$S \\ 
    Symbol time: $S$ & $\frac{1}{9\cdot 10^6}$ s/c.u. & $P_\mathrm{rx}$&0.3 W$\cdot$S\\
    Noise variance & $10^{-20}$ J/c.u.&$\eta$&0.3\\\hline
  \end{tabular}
\label{table:parameters}
\end{table}\vspace{-0.2cm} 

\begin{figure}[h!]
\begin{centering}
\includegraphics [height=0.26\textwidth]{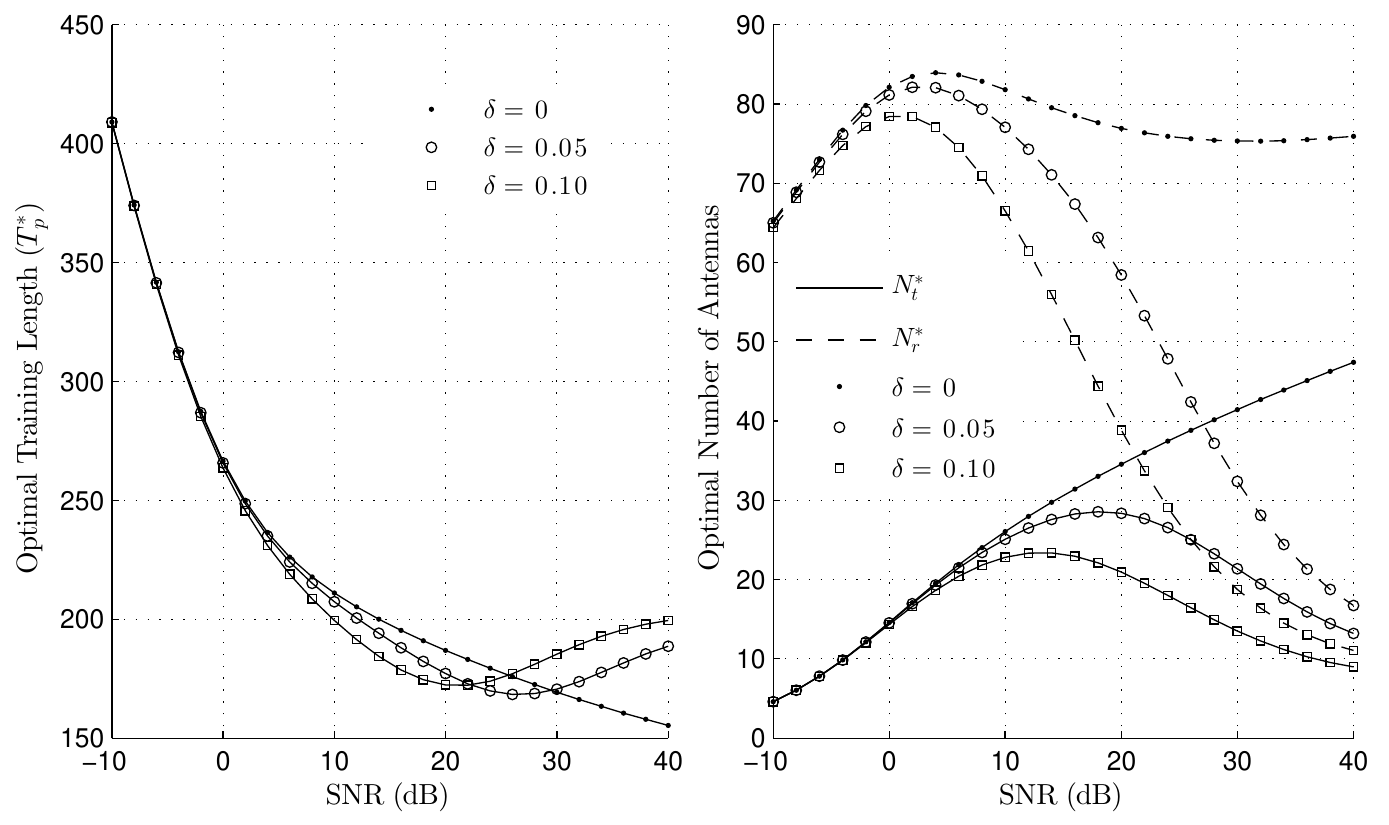}
\caption{Optimal training length and antenna configurations for different levels of RTRI.}\label{fig:Tp_and_NrNt}
\end{centering}
\end{figure}

Figure \ref{fig:Tp_and_NrNt} illustrates the optimal training length, and numbers of transmit and receiver antennas which maximize the energy efficiency. Our simulation parameters are inspired by \cite{Auer2011EE_TWCOM, Emil2014EE_TWCOM} and summarized in Table \ref{table:parameters}. We choose the threshold $\overline{\mathrm{EE}}_\mathrm{th} = 10^{-10}$ bit/Joule. For systems with and without hardware impairments, we can clearly see the difference in the tradeoff between training and data transmission from moderate to high SNR values. We can interpret this  by observing the energy efficiency in this specific SNR regime 
\vspace{-0.2cm}
\begin{equation}\label{eq:EE_2}
\overline{\mathrm{EE}} \approx \frac{\left(1\!-\!\frac{T_p}{N_t}\right)N_t\log_2\left(\!1\!+\!\frac{\beta-1}{\delta^2(\beta-1) +\frac{N_t}{T_p}(\delta^2+1\!/\!\rho)+1\!/\!\rho}\right)}{P_\mathrm{total}(N_r, N_t, \rho)},
\end{equation}
where we have ignored the minor terms $\delta^4$, $1\!/\!\rho^2$ and $2\delta^2/\rho$.
For systems with ideal hardware, which corresponds to $\delta = 0$, increasing $T_p$ cannot provide any gain in the SINR (the function inside the $\log_2(\cdot)$ term) in the high SNR regime, but monotonically decreases the ratio of coherence block assigned to data transmission. Thus, only $T_p = N_t$ c.u.s are needed for training as $\rho\rightarrow\infty$. For systems with RTRI, however, increasing $T_p$  always increases the channel estimation accuracy and, thus, the SINR, despite of the SNR value. The optimal training length $T_p^*$, obviously, depends on the SNR and the level of RTRI. In general, for systems with RTRI considered herein, to maximize the energy efficiency  less training is needed at moderate SNR values, while more training is required at high SNR values, compared to ideal hardware systems.%These important observations provide a guideline for the fundamental tradeoff between training and data transmission. 

The right-hand side subplot of Fig.~\ref{fig:Tp_and_NrNt} demonstrates the optimal numbers of antennas. For ideal systems, the optimal number of transmit antennas,  $N_t^*$,  increases monotonically as SNR increases, and the optimal number of receive antennas, $N_r^*$, is around 70 and does not vary too much, apart from at unrealistically high SNR values which are not depicted in this figure.\footnote{Note that the result varies for different power consumption parameters.} For systems with RTRI, however, much less transmit and receive antennas are needed from moderate to high SNR values. This implies that, increasing the spectral efficiency by adding more antennas can lead to a penalty in the energy efficiency for systems with RTRI. This can be understood by examining (\ref{eq:EE_2}): due to the term $\delta^2(\beta-1)$, the SINR of a system with RTRI is not increased with $N_r$ (or $\beta$) as fast as that of a system with ideal hardware, while the total power consumption still increases linearly with $N_r$. Therefore, much less numbers of receive antennas are needed to achieve the maximal energy efficiency. On the transmit side, adding more antennas in systems with RTRI at high SNR values decreases the SINR much more than in ideal systems; consequently, we should use fewer transmit antennas in hardware-impaired systems to guarantee a maximal energy efficiency.

\begin{figure}[h!]
\begin{centering}
\includegraphics [height=0.23\textwidth]{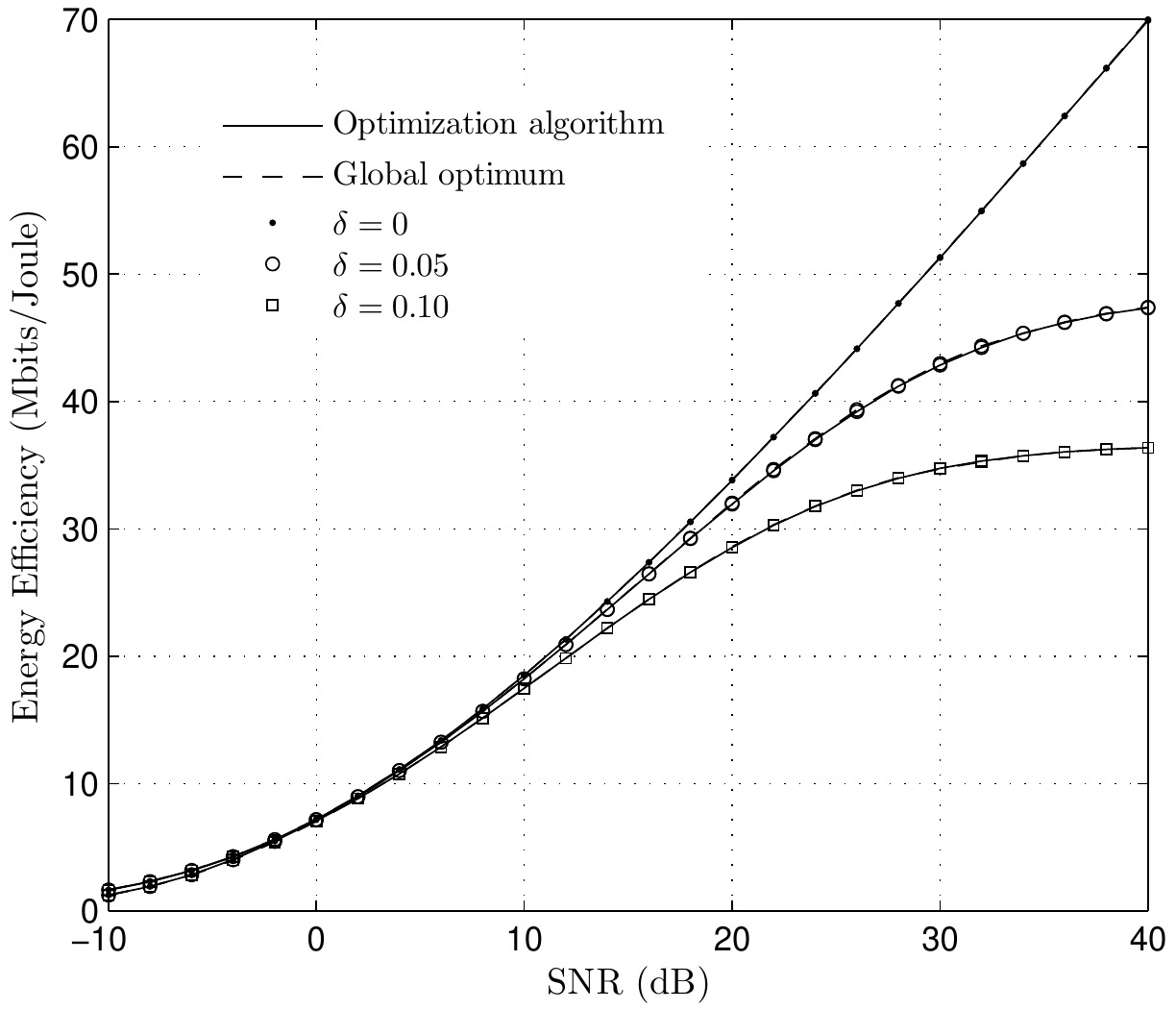}
\caption{Comparison of the global optimum and iterative sequentially optimized energy efficiency.}\label{fig:EEopt}
\end{centering}
\end{figure}

In Fig. \ref{fig:EEopt}, we compare the globally optimal energy efficiency (achieved by extensive search) and the one achieved by our iterative sequential optimization. We see excellent match between these two algorithms, which validates the effectiveness of the proposed optimization scheme. We observe that RTRI decrease the energy efficiency significantly in the high SNR regime. Yet, for low SNR values, there is no substantial difference between ideal  systems and systems with RTRI.
\vspace{-0.3cm}
\vspace{-0.5cm}
\vspace{-0.9cm}
\section{Conclusion}
In this paper, we analyzed the performance of point-to-point training-based large-scale MIMO systems with residual transmit RF impairments. In particular, we derived  deterministic equivalents of the SINR, spectral efficiency, and energy efficiency of ZF receivers. These asymptotic expressions were shown to provide very accurate approximations even for small number of antennas. Furthermore, we optimized the energy efficiency with respect to the training length, and the number of antennas, through an iterative sequential algorithm. We found that RTRI have significant impact on these parameters. More specifically, systems with RTRI required less training in the moderate SNR regime, while needed more training in the high SNR regime. Thus, a good tradeoff between training and data transmission is crucial to obtain the maximal energy efficiency. In addition, the optimal numbers of transmit and receive antennas were much smaller than those of systems with ideal hardware. Finally, we demonstrated an energy efficiency penalty caused by RTRI in the moderate and high SNR regime. This work provided guidelines for designing the optimal training length and number of antennas in point-to-point large-scale MIMO systems with hardware impairments.

\section*{Acknowledgment}
The work of X.~Zhang, M.~Matthaiou and M. Coldrey has been supported in part by the Swedish Governmental Agency for Innovation Systems (VINNOVA) within the VINN Excellence Center Chase. The work of E.~Bj{\"o}rnson was supported by the International Postdoc Grant 2012-228 from the Swedish Research Council.

\appendices

\bibliographystyle{IEEEtran}
\bibliography{IEEEabrv,Reference}

\end{document}